\documentclass[runningheads,a4paper]{llncs}

\usepackage{amssymb}
\usepackage{graphicx}
\usepackage{booktabs}
\usepackage{multirow}

\usepackage{url}
\urldef{\mailsa}\path{naldi@disp.uniroma2.it}
\newcommand{\keywords}[1]{\par\addvspace\baselineskip
\noindent\keywordname\enspace\ignorespaces#1}

\begin{document}

\mainmatter  

\title{A note on "The Need for End-to-End Evaluation of Cloud Availability"}

\titlerunning{Cloud availability}

%
%
\author{Maurizio Naldi\inst{1}}
\institute{Dept. of Computer Science and Civil Engineering, Universit\`{a} di Roma Tor Vergata, Roma, Italy,\\
\email{naldi@disp.uniroma2.it},\\ WWW home page:
\texttt{http://www.maurizionaldi.it}
}

\authorrunning{Maurizio Naldi}


%
%

\toctitle{Lecture Notes in Computer Science}
\tocauthor{Authors' Instructions}
\maketitle

\begin{abstract}
Cloud availability is a major performance parameter for cloud platforms, but there are very few measurements on commercial platforms, and most of them rely on outage reports as appeared on specialized sites, providers' dashboards, or the general press. A paper recently presented at the PAM 2014 conference by Hu et alii reports the results of a measurement campaign. In this note, the results of that paper are summarized, highlighting sources of inaccuracy and some possible improvements. In particular, the use of a low probing frequency could lead to non detection of short outages, as well as to an inaccurate estimation of the outage duration statistics. Overcoming this lack of accuracy is relevant to properly assess SLA violations and establish the basis for insurance claims.

\end{abstract}

\section{Introduction}
Cloud availability is a major performance parameter for cloud platforms. When an individual or a company switch to the cloud, they wish the platform to be at least as available as their in-house infrastructure. For that reason, availability is always present among the parameters to be monitored in cloud monitoring systems \cite{montes2013} and to be considered in cloud platform assessment systems\cite{Li-GRID-2012}\cite{CalheirosSPE995}.
It is typically considered for inclusion in SLAs \cite{CuomoJGC2013}\cite{Emeakaroha-FGCS2012}\cite{gillam2012}, and its values are boasted of on all major commercial cloud platforms. 

While a model for the prediction of cloud service reliability has been proposed in \cite{dai2009}, not many efforts are present in the literature to investigate the availability actually offered on commercial platforms. For example, availability is not considered in the comparison carried out in \cite{li2010cloudcmp}. In the description of a  major commercial platform, Microsoft Azure, provided in \cite{calder2011windows}, though a high availability is claimed right in the title, no figures are provided for the expected availability. A notable exception is represented by the study published in 2010 by Ford et alii, who performed a wide set of measures on Google's storage infrastructure, though the results concern single nodes rather than the cloud platform as a whole \cite{ford2010}, i.e., they reflect a cloud-centric point of view. Though the authors did not provide many details on the measurement method, they appear to employ the straightforward approach of pinging the storage nodes and recording the presence of the echo. The reliability of servers in cloud infrastructure has been analysed in \cite{vishwanath2010}. Previous efforts on cloud availability characterization from a user-centric point of view fall instead in the category of indirect methods, relying on reported data rather than actual measurements.  Data from cloud provider status dashboards and press releases have been collected and analysed in \cite{IWGCR2013}; such data have also been reported on the International Working Group on Cloud Computing Resiliency (IWGCR) website (\url{www.iwgcr.org}). Again, data from the IWGCR and from Cloutage (\url{cloutage.org}), an organization founded by the Open Security Foundation, have been analysed in \cite{Naldi-DRCN13}, where a Generalized Pareto model has been provided for the tail of the outage duration distribution. Finally, in \cite{Motoyama2010} Twitter messages have been collected and analysed to infer the availability of Internet services (mostly cloud-based).

In the latest edition of the Passive and Active Measurements Conference (PAM 2014), a paper has been presented on end-to-end availability measurements \cite{Hu-PAM2014}. The paper represents a new attempt to provide third party measurements of the availability of commercial cloud providers and represents a step forward with respect to the previous efforts, which were mainly based on indirect reports. In this note, we review its main results in Section \ref{results} and analyse possible ways to improve that analysis in Section \ref{issues}.

\section{What the paper accomplishes}
\label{results}
In this section we review the main results reported in the paper by Hu et alii.

Hu et alii have compared the performances of two approaches to measure cloud availability, both for computing (creation and use of virtual machines) and storage applications. The two approaches make use respectively of the protocols ICMP and HTTP, hence they are dubbed by the authors themselves as network and application level measurements respectively. The test is claimed to be conducted on three cloud providers (Amazon Simple Storage Service S3, Microsoft Azure, and Google Cloud Storage), though VMs are tested on Amazon only and results for storage are provided for two providers only (Amazon and Google). The results are anyway very interesting in the current deserted panorama of third-party measurements of cloud platforms. The measurements have been obtained during two separate measurement campaigns, lasting respectively 33 and 75 days. The tests were conducted from a large number of vantage points in the PlanetLab network \cite{chun2003planetlab}.

The main aim of the research work was to compare ICMP and HTTP from the accuracy viewpoint, where the accuracy (in the absence of a reference, which should be based on measurements conducted inside the cloud) is evaluated as the capability of filtering out non-cloud failures (e.g., misrouting events and packet losses on the way to/from the cloud). 

In the case of storage testing, the ICMP approach consists in sending a probe message to the front-end server (namely the hostname in the URL of the stored object) and counting the echoes as a measure of success following the same approach as in \cite{ford2010}, while in the HTTP approach the test consists in retrieving a file previously stored on the cloud under test. The HTTP approach allows to test the whole cloud service, while ICMP just tests the functionality of the single front-end server. Both type of tests allow for a number of retries in the case of a failure, on the assumption that short-lived network failures are so removed from the count (this equals to assuming that cloud failures are not short-lived).  

The main result is that ICMP tests are affected by network-related failures more than HTTP, and ICMP can err in either ways, sometimes overestimating the actual failure rate, while underestimating at other times. The conclusion is that the HTTP approach is to be preferred to obtain a measure of the actual cloud availability. Though the authors do not provide explicit figures for the availability estimate after the treatment of retries, these can be obtained through the figures provided for the availability estimate considering the first try only and the subsequent conditional failure probability (though the latter are provided through a graph, rather than with precise figures). The estimated availability based on first tries only is shown in Table \ref{table:estimates} for storage tests: it is somewhat midway between 2 and 3 nines.

\section{Pending issues}
\label{issues}
While in Section \ref{results} we have surveyed the main results reported in the paper, in this section we report some critical issues that deserve further work.

A first limitation is the number of cloud platforms under analysis, which is just 1 for VMs and 2 for storage (Amazon Simple Storage Service S3 and Google Cloud Storage), though the authors state that they are actually considering 3 providers. This is certainly a step forward with respect to the current unavailability of third-party measurements, but it is still not enough to obtain a panorama of the performances we can expect of a cloud.

\begin{table}
\begin{center}
\begin{tabular}{lcc}
\toprule
Platform & Prob [1st try fails] & Availability [\%]\\
\midrule
Amazon/storage & 0.00435 & 99.565\\
Google/storage & 0.00217 & 99.783\\
\bottomrule\\
\end{tabular}
\caption{Availability estimates}
\label{table:estimates}
\end{center}\end{table}

If we turn to the measurement methodology, we examine first the use of retries. The authors report the probability that the first probing message fails, which can be considered as an estimate of the availability, as shown in Table \ref{table:estimates}, where we report the results for the HTTP-based approach only, since it provides a better estimate than ICMP.  In an attempt to filter out outages that are not to be debited to the cloud platform, Hu et alii adopt a retry mechanism after each failed attempt, with a maximum of 9 attempts. If we indicate by $y_{i}$ the number of $i$-th attempts and by $x_{i}$ the number of successful $i$-th attempts, the following equality holds true for a maximum of $n$ attempts.
\begin{equation}
y_{i} = y_{1}-\sum_{j=1}^{i-1}x_{j} \qquad i=2, 3, \ldots, n,
\end{equation}
In the paper, they adopt as an estimate of the availability the quantity
\begin{equation}
\label{paperest}
p^{*} = \frac{\sum_{j=1}^{n}x_{j} }{y_{1}}.
\end{equation}

Though that estimate may help getting rid of very short-lived failures that are due to the in-between infrastructure, it also gets rid of short-lived failures that are actually due to the cloud itself. Hence, the paper makes the implicit assumption that there are very short-lived (whose duration is of the order of seconds) network failures, while there are no such cloud failures. 

On the other hand, the estimate of Equation (\ref{paperest}) may represent an overestimate of the cloud availability, for which a better estimate could be
\begin{equation}
p_{1} = \frac{x_{1}}{y_{1}}
\end{equation}
or, assuming that the response to successive attempts is independent of past failures
\begin{equation}
p_{n} = \frac{\sum_{i=1}^{n}x_{i}}{\sum_{i=1}^{n}y_{i}}.
\end{equation}
In fact, the method of retries, if pushed too far, would lead to certainly wrong results, since $\lim_{n\rightarrow\infty}p^{*}=1$. In the absence of network failures, the introduction of further attempts after each failure leads to estimate the geometric probability $\sum_{i=1}^{n}p(1-p)^{i-1}$ when the availability is $p$. The resulting overestimation factor is
\begin{equation}
f = \sum_{i=1}^{n}(1-p)^{i-1} = \sum_{i=0}^{n-1}(1-p)^{i} = \frac{1-(1-p)^{n}}{p}.
\end{equation}
For a $k$-nines availability, we have
\begin{equation}
f = \frac{1-10^{-kn}}{1-10^{-k}}.
\end{equation}
This factor is anyway very close to 1. For $n=9$ attempts, we have $f\simeq 1.001$ for a 3-nines availability and $f\simeq 1.01$ for a 2-nines  availability. Though we have considered the responses to successive retries independent of each other, while there may be a strong correlation, even such a small factor may have a significant impact when the availability is high. In fact, a 1.01 factor would transform a 2 nines availability into a 4 nines one. The retry mechanism should therefore be considered as safe when the assumption about the presence of very short-lived failures has been validated.

However, the paper does not investigate the respondence of present clouds to their SLA claims and users' requirements. As can be seen in Table \ref{table:estimates}, the estimate $p_{1}$ (which can be considered as a lower bound for the availability estimate, in the light of the previous discussion) lies between 2 and 3 nines, falling short of what most cloud providers claim: in the survey reported in \cite{CasalicchioS13}, 15 providers out of 17 declared at least 99.9\% availability, with 12 providers declared 100\% availability. If we had the exact number of tries, we could compute confidence intervals for the availability estimate and perform a statistical test to decide whether to reject or not the null hypothesis that the provider complies with its availability-related SLA claims. Unfortunately the authors do not release the exact figure, but we can try and infer it from other data the authors provide. Considering the declared frequency of tries (1 every either 10 or 11 minutes), and assuming that tries are carried out continuously during both measurement campaign periods and from each of the vantage points, we obtain an overall number of tries equal to $y_{1}=639478$ (see Table \ref{table:datmis} for the details)
\begin{table}
\begin{tabular}{lcccc}
\toprule
Period start & Repetition interval [minutes] & Duration [days] & Vantage points & No. of tries\\
\midrule
11 March 2013 & 10 & 33 & 23 & 109296\\
18 June 2013 & 11 & 75 & 54 & 530182\\
\midrule
Total & & & & 639478\\
\bottomrule\\
\end{tabular}
\caption{Measurement campaign data}
\label{table:datmis}
\end{table}
This leads us to an approximate standard deviation $\sigma$ of the availability estimated that is $8.2\cdot 10^{-5}$ for Amazon and $5.8\cdot 10^{.5}$ for Google, by using the formula 
\begin{equation}
\sigma \simeq \sqrt{\frac{p_{1}(1-p_{1})}{y_{1}} }.
\end{equation}
If the estimated availability were $p_{1}$, such a tight confidence interval would lead us to reject the hypothesis that the actual availability is a triple nine or better.
 
\begin{figure}
\begin{center}
\includegraphics[width=0.5\columnwidth]{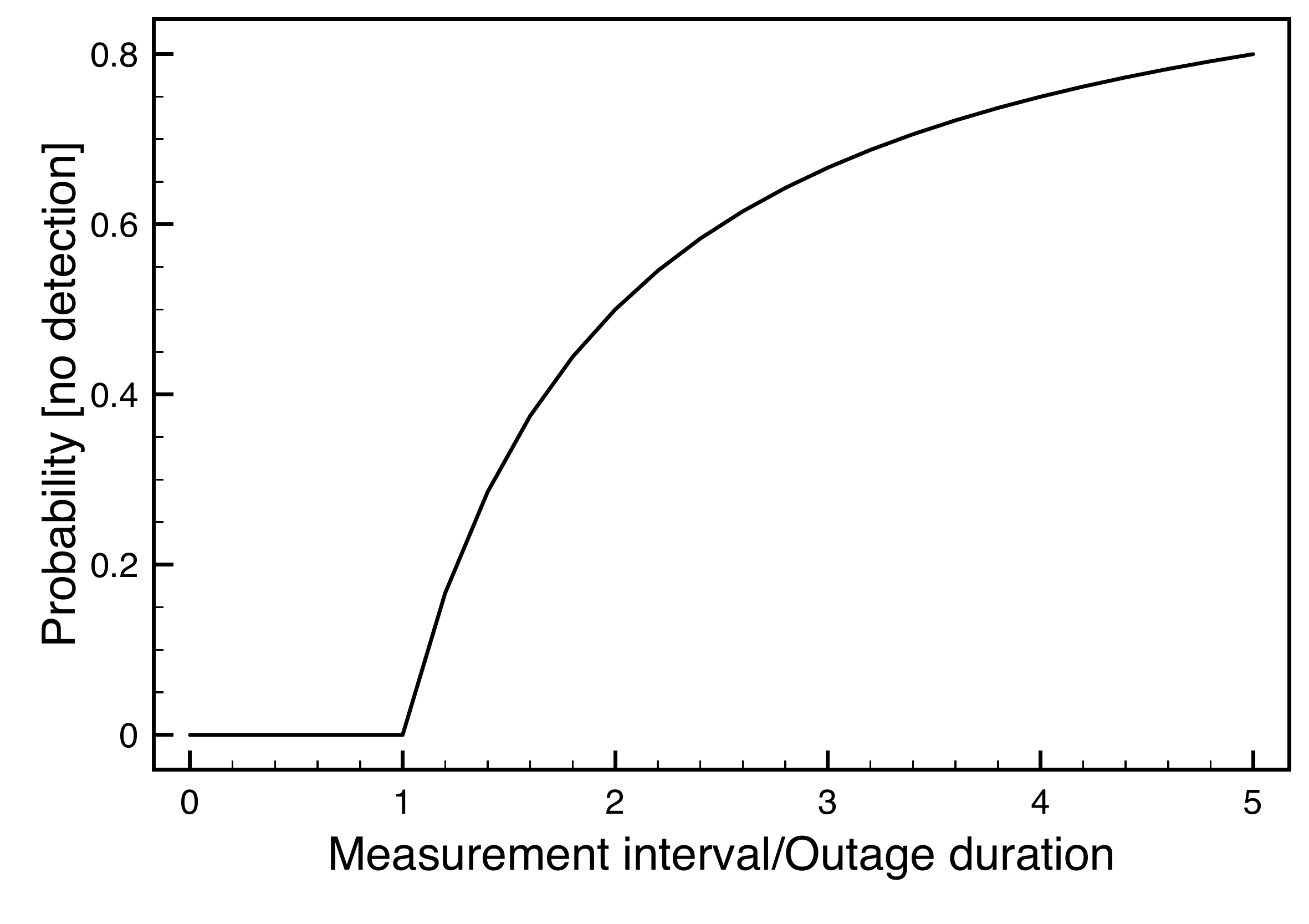}
\caption{Probability of an outage going undetected}
\label{fig:nodet} 
\end{center}
\end{figure}

Another point to consider is the observation of failures lasting for more than a few seconds but shorter than the measurement interval. As already reported in Table \ref{table:datmis}, the frequency of probing messages (excluding the rare retries) is either 10 or 11 minutes. That means that outages of duration lower than that interval are severely censored, since they may not be detected at all. In fact, if we mark the occurrence of the measurement timepoint preceding the failure as time 0, so that the next measurement takes place at the time $T$ (e.g., 10 or 11 minutes in this case), the failure will take place at a random time $X$ such that $0\le X \le T$. If we consider $X$ to be uniformly distributed, the failure will not be detected if the recovery is achieved before the next measurement interval. If the outage duration is $L$, that condition can be expressed as $X + L < T$. The probability that the outage goes undetected is then
\begin{equation}
P_\mathrm{nodet}=\mathbb{P}[X + L < T] = \mathbb{P}\left[\frac{X}{T}<1-\frac{L}{T}\right] = \left\{ \begin{array}{ll}
 0 & \textrm{if $T\le L$}\\
 1-\frac{L}{T} & \textrm{if $T>L$}
  \end{array} \right.
\end{equation}
The resulting no-detection probability is shown in \figurename~\ref{fig:nodet}.

Finally, the use of such a long meauserement interval makes it difficult to obtain an accurate  statistical distribution of outage durations. In turn, it becomes difficult to compute the extent of SLA violations and the amount of possible compensations or insurance claims \cite{Naldi-Fedcsis11}\cite{Naldi-ICQT11}\cite{CholdaFHKNN13}. In fact,  of the three measures envisaged to be used in an insurance policy for network failures in \cite{NaldiN11} (but applicable straightforwardly to the case of clouds) just one is attempted in the paper:
\begin{itemize}
\item Number of failures;
\item Number of outages lasting more that a prescribed threshold (long outages);
\item Cumulative outage duration.
\end{itemize}
The cumulative outage duration is equal to the overall unavailability, which the paper provides, but the number of failures is severely distorted, since short-lived outages may go undetected, and the number of long outages may be underestimated as well unless the threshold is longer than the measurement interval (10 or 11 minutes in this case). 

\section{Conclusions}
We have reviewed the paper "The Need for End-to-End Evaluation of Cloud Availability" by Hu et alii, recently presented at the 2014  edition of the Passive and Active Measurement conference (PAM).

The paper reports the results of two measurement campaigns to compare two methods of estimating cloud availability, based on the use of the ICMP and HTTP protocol respectively, while the literature shows very few observations of cloud providers' performances, mostly based on an indirect approach (using reported data rather than direct measurements).

The main result is that the HTTP-based approach is more accurate and allows to filter off some network-related failures.

Among the cons, a more careful discrimination of network-related failures is to be carried out for a proper assessment of the estimation methods. The current approach and its choice of probing frequency could censor short-lived cloud outages. In addition, there is no attempt to estimate the statistical distribution of outage durations, which is needed for a proper assessment of the probability of SLA violations and the ensuing damage assessment.



\begin{thebibliography}{10}
\providecommand{\url}[1]{\texttt{#1}}
\providecommand{\urlprefix}{URL }

\bibitem{calder2011windows}
Calder, B., Wang, J., Ogus, A., Nilakantan, N., Skjolsvold, A., McKelvie, S.,
  Xu, Y., Srivastav, S., Wu, J., Simitci, H., et~al.: Windows azure storage: a
  highly available cloud storage service with strong consistency. In:
  Proceedings of the Twenty-Third ACM Symposium on Operating Systems
  Principles. pp. 143--157. ACM (2011)

\bibitem{CalheirosSPE995}
Calheiros, R.N., Ranjan, R., Beloglazov, A., De~Rose, C.A.F., Buyya, R.:
  Cloudsim: a toolkit for modeling and simulation of cloud computing
  environments and evaluation of resource provisioning algorithms. Software:
  Practice and Experience  41(1),  23--50 (2011)

\bibitem{CasalicchioS13}
Casalicchio, E., Silvestri, L.: Mechanisms for {SLA} provisioning in
  cloud-based service providers. Computer Networks  57(3),  795--810 (2013)

\bibitem{IWGCR2013}
C\'{e}rin, C., Coti, C., Delort, P., Diaz, F., Gagnaire, M., Gaumer, Q.,
  Guillaume, N., Lous, J.L., Lubiarz, S., Raffaelli, J.L., Shiozaki, K.,
  Schauer, H., Smets, J.P., S\'{e}guin, L., Ville, A.: Downtime statistics of
  current cloud solutions. Available at
  \url{http://iwgcr.org/wp-content/uploads/2013/06/IWGCR-Paris.Ranking-003.2-en.pdf}
  (June 2013)

\bibitem{CholdaFHKNN13}
Cholda, P., F{\o}lstad, E.L., Helvik, B.E., Kuusela, P., Naldi, M., Norros, I.:
  Towards risk-aware communications networking. Rel. Eng. {\&} Sys. Safety
  109,  160--174 (2013)

\bibitem{chun2003planetlab}
Chun, B., Culler, D., Roscoe, T., Bavier, A., Peterson, L., Wawrzoniak, M.,
  Bowman, M.: Planetlab: an overlay testbed for broad-coverage services. ACM
  SIGCOMM Computer Communication Review  33(3),  3--12 (2003)

\bibitem{CuomoJGC2013}
Cuomo, A., Di~Modica, G., Distefano, S., Puliafito, A., Rak, M., Tomarchio, O.,
  Venticinque, S., Villano, U.: An sla-based broker for cloud infrastructures.
  Journal of Grid Computing  11(1),  1--25 (2013),
  \url{http://dx.doi.org/10.1007/s10723-012-9241-4}

\bibitem{dai2009}
Dai, Y.S., Yang, B., Dongarra, J., Zhang, G.: Cloud service reliability:
  Modeling and analysis. In: 15th IEEE Pacific Rim International Symposium on
  Dependable Computing (2009)

\bibitem{Emeakaroha-FGCS2012}
Emeakaroha, V.C., Netto, M.A., Calheiros, R.N., Brandic, I., Buyya, R., Rose,
  C.A.D.: Towards autonomic detection of \{SLA\} violations in cloud
  infrastructures. Future Generation Computer Systems  28(7),  1017 -- 1029
  (2012), special section: Quality of Service in Grid and Cloud Computing

\bibitem{ford2010}
Ford, D., Labelle, F., Popovici, F.I., Stokely, M., Truong, V.A., Barroso, L.,
  Grimes, C., Quinlan, S.: Availability in globally distributed storage
  systems. In: 9th USENIX Symposium on Operating Systems Design and
  Implementation, OSDI 2010. pp. 61--74. Vancouver, BC, Canada (October 4-6,
  2010)

\bibitem{gillam2012}
Gillam, L., Li, B., O'Loughlin, J.: Adding cloud performance to service level
  agreements. In: CLOSER. pp. 621--630 (2012)

\bibitem{Hu-PAM2014}
Hu, Z., Zhu, L., Ardi, C., Katz-Bassett, E., Madhyastha, H., Heidemann, J., Yu,
  M.: The need for end-to-end evaluation of cloud availability. In: Faloutsos,
  M., Kuzmanovic, A. (eds.) Passive and Active Measurement, Lecture Notes in
  Computer Science, vol. 8362, pp. 119--130. Springer International Publishing
  (2014), \url{http://dx.doi.org/10.1007/978-3-319-04918-2_12}

\bibitem{li2010cloudcmp}
Li, A., Yang, X., Kandula, S., Zhang, M.: Cloudcmp: comparing public cloud
  providers. In: Proceedings of the 10th ACM SIGCOMM conference on Internet
  measurement. pp. 1--14 (2010)

\bibitem{Li-GRID-2012}
Li, Z., O'Brien, L., Zhang, H., Cai, R.: On a catalogue of metrics for
  evaluating commercial cloud services. In: Grid Computing (GRID), 2012
  ACM/IEEE 13th International Conference on. pp. 164--173 (Sept 2012)

\bibitem{NaldiN11}
Mastroeni, L., Naldi, M.: Network protection through insurance: Premium
  computation for the on-off service model. In: 8th International Workshop on
  the Design of Reliable Communication Networks DRCN, Krakow, Poland. pp.
  46--53 (10-12 October 2011)

\bibitem{Naldi-ICQT11}
Mastroeni, L., Naldi, M.: Compensation policies and risk in service level
  agreements: A value-at-risk approach under the on-off service model. In:
  Economics of Converged, Internet-Based Networks - 7th International Workshop
  on Internet Charging and QoS Technologies, ICQT 2011, Paris, France. Lecture
  Notes in Computer Science, vol. 6995, pp. 2--13. Springer (October 24, 2011)

\bibitem{montes2013}
Montes, J., S{\'a}nchez, A., Memishi, B., P{\'e}rez, M.S., Antoniu, G.: Gmone:
  A complete approach to cloud monitoring. Future Generation Computer Systems
  29(8),  2026--2040 (2013)

\bibitem{Motoyama2010}
Motoyama, M., Meeder, B., Levchenko, K., Voelker, G.M., Savage, S.: Measuring
  online service availability using twitter. In: Proceedings of the 3rd
  Wonference on Online Social Networks. pp. 13--13. WOSN'10, USENIX
  Association, Berkeley, CA, USA (2010)

\bibitem{Naldi-DRCN13}
Naldi, M.: The availability of cloud-based services: Is it living up to its
  promise? In: 9th International Conference on the Design of Reliable
  Communication Networks, DRCN 2013, Budapest, Hungary. pp. 282--289 (March
  4-7, 2013)

\bibitem{Naldi-Fedcsis11}
Naldi, M., Mastroeni, L.: Violation of service availability targets in service
  level agreements. In: Federated Conference on Computer Science and
  Information Systems - FedCSIS 2011, Szczecin, Poland. pp. 537--540 (18-21
  September 2011)

\bibitem{vishwanath2010}
Vishwanath, K.V., Nagappan, N.: Characterizing cloud computing hardware
  reliability. In: Proceedings of the 1st ACM symposium on Cloud computing
  SoCC. pp. 193--204. ACM (2010)

\end{thebibliography}

\end{document}